\documentclass[aps,prb,reprint,showpacs]{revtex4-1}

\usepackage{amsmath}
\usepackage{amssymb}
\usepackage{bm}
\usepackage{color}
\usepackage{graphicx}
\usepackage[latin1]{inputenc}
\usepackage[english]{babel}
\begin{document}


\title{Temperature dependence of linewidth in nano-contact based spin torque oscillators: effect of multiple oscillatory modes}
\author{P.~K.~Muduli}
\email{pranaba.muduli@physics.gu.se}
\affiliation{Department of Physics, University of Gothenburg, 41296 Gothenburg, Sweden}
\affiliation{Department of Physics, Indian Institute of Technology Delhi, New Delhi, 110016, India}

\author{O.~G.~Heinonen}
\affiliation{Materials Science Division, Argonne National Laboratory, Lemont, IL 60439, USA}
\affiliation{Department of Physics and Astronomy, Northwestern University, 2145 Sheridan Rd., Evanston, IL 60208-3112}

\author{Johan~\AA kerman}
\affiliation{Physics Department, University of Gothenburg, 41296 Gothenburg, Sweden}
\affiliation{Materials Physics, School of ICT, KTH-Royal Institute of Technology,
Electrum 229, 164 40 Kista, Sweden}

\begin{abstract}
We discuss the effect of mode transitions on the current ($I$) and temperature ($T$) dependent linewidth ($\Delta f$) in nanocontact based spin torque oscillators (STOs). At constant $I$, $\Delta f$  exhibits an anomalous temperature dependence near the mode transitions; $\Delta f$ may either increase or decrease with $T$ depending on the position w.r.t. the mode transition.
We show that the behavior of $\Delta f$ as a function of $I$ can be fitted by the single mode analytical theory of STOs, even though there are two modes present near the mode transition, if the nonlinear amplification is determined directly from the experiment. Using a recently developed theory of two \emph{coupled} modes, we show that the linewidth near mode transition can be described by an ``effective" single-oscillator theory with an enhanced nonlinear amplification that carries additional temperature dependence, which thus qualitatively explain the experimental results.
\end{abstract}
 \maketitle


\section{INTRODUCTION}
A spin-polarized current traversing a thin magnetic layer can exert a significant torque on the magnetization through the spin transfer torque (STT) effect.~\cite{slonczewski1996jmmm,berger1996prb,tsoi1998prl,tsoi2000nt,kiselev2003nt,ralph2008jmmm,sun2008jmmm} The effect can be described as negative damping, linearly proportional to the spin-polarized current, which at a certain threshold 
can overcome the natural Gilbert damping 
in the magnetic layer, allowing for coherent, large amplitude, excitation of spin waves. If the magnetic layer is part of a structure with magnetoresistance, such as a spin valve (SV) or a magnetic tunnel junction (MTJ), the excited spin waves can be used to generate a current- and field-tunable microwave voltage signal; the resulting device is commonly called a spin torque oscillator (STO).~\cite{silva2008jmmm} Interest in STOs for microwave applications is steadily increasing, due to their attractive combination of very large frequency tuning ranges,~\cite{rippard2004prb,bonetti2009apl,muduli2011jap} efficient  spin-wave  emission  in  magnonic
devices,~\cite{bonetti2010prl,demidov2010ntm,madami2011nn} very high modulation rates,~\cite{pufall2005apl,manfrini2009apl,muduli2010prb,muduli2011if,pogoryelov2011apl,pogoryelov2011apldc,manfrini2011jap,muduli2011ieeem,muduli2011aip} sub-micron footprints,~\cite{vincent2009ieeejssc} and straightforward integration with semiconductor technology using the same processes as magnetoresistive random access memory.~\cite{engel2005ieeemag, akerman2005sc}

A minimal spectral linewidth, $\Delta f$, of the microwave signal is highly desirable for applications. While a number of recent experimental studies have addressed the temperature dependence of $\Delta f$ in nanopillar STOs ~\cite{sankey2005prb, mistral2006apl, georges2009prb, boone2009prb,bortolotti2012apl,sierra2012apl}
the study of the temperature dependent linewidth in nanocontact STOs is limited to a recent work by Schneider \emph{et. al.}~\cite{schneider2009prb} The theory of the origin of STO linewidths and their temperature dependence is now well established for single spin-wave modes.~\cite{kim2006prb,kim2008prl1,kim2008prl, tiberkevich2008prb, slavin2009ieeem, silva2010ieeemag} A key result is the strong impact that limited amplitude noise can have on the STO phase noise, via the strong amplitude-phase coupling. Gaussian (white) amplitude noise is transformed into colored phase noise, and the intrinsic Lorentzian line shape expected for an auto-oscillator with zero amplitude-phase coupling changes into a convolution of Lorentzian and Gaussian line shapes.~\cite{keller2010prb} The coupling also leads to a substantial enhancement, or amplification, of the thermal broadening, and can also lead to asymmetric line shapes near threshold.~\cite{kim2008prl} The degree of coloring should also change with temperature, leading to a crossover from a linear temperature dependence of $\Delta f$ at low temperature, to a square root dependence at high temperature.~\cite{tiberkevich2008prb}

All temperature dependent studies to date show temperature regions with unexpected behavior. In Ref.~\onlinecite{georges2009prb}, $\Delta f$ in the subthreshold regime narrows by a factor of 6, from 1.2 GHz to 200 MHz, when the temperature is raised from 20 K to 140 K. In Ref.~\onlinecite{mistral2006apl}, the slope of the temperature dependence even changes sign multiple times as a function of drive current, and is close to zero at the smallest $\Delta f$. In Ref.~\onlinecite{sankey2005prb}, $\Delta f$ increases exponentially above a certain temperature; the concept of mode hopping was introduced to explain and model this dependence. The origin of these rather complex temperature dependencies is yet to be explained. More recently, a linear behavior of linewidth is observed for a certain range of temperature in magnetic tunnel junction based STOs.~\cite{bortolotti2012apl,sierra2012apl} A saturation of linewidth is observed in both these studies for temperature below 100 K, which is not explained by the existing theories. In addition, the temperature dependence of power restoration rate observed in Ref~\onlinecite{sierra2012apl} can not be explained by the single mode theory.~\cite{slavin2009ieeem} Thus the details of the temperature dependence of linewidth in STOs is far from being understood.

In this work, we present a detailed study of the temperature-dependent linewidth in nanocontact STOs. While all measurements were carried out at current and magnetic field values where only propagating spin waves were generated,~\cite{bonetti2010prl,bonetti2012prb} we found a large number of mode transitions as a function of current (at a fixed temperature $T$) \emph{and} temperature (at a fixed current $I$). The measured linewidth is highly nonmonotonic both as a function of current and of temperature, with large enhancements at currents or temperatures where mode transitions occurred. We show that the linewidth is very well fitted by the single oscillator theory~\cite{tiberkevich2008prb, slavin2009ieeem}, if the so-called amplification factor
is obtained directly from measurements. While this agreement is similar to that of Refs.~\onlinecite{georges2009prb,pogoryelov2011apl}, we find the temperature dependence of the linewidth does not agree with that obtained directly from calculations using the nonlinear single-oscillator
theory~\cite{tiberkevich2008prb,slavin2009ieeem}, from which
typically a linear dependence on $T$ is obtained for the systems under study here. These observations indicate that the central mechanism for
linewidth broadening in
nonlinear single-oscillator theory applies here, too: The linewidth is driven by phase noise amplified by the coupling through the
nonlinear frequency shift to power amplitude
fluctuations. However, our results indicate that this coupling may itself have a nontrivial temperature (and current) dependence, especially
near mode transitions. We will here show that extending the nonlinear single-oscillator theory to include two coupled modes~\cite{muduli2012prl} leads
to additional couplings between the phase and power fluctuations. Under some simplifying assumptions,
these couplings lead to a changed power restoration rate and the final result for the linewidth looks
very much like that from the nonlinear single-oscillator theory\cite{tiberkevich2008prb, slavin2009ieeem}, but with an
enhanced nonlinear amplification that carries additional temperature dependence. This explains qualitatively the observed temperature dependence
of the linewidth near mode transitions.

\section{EXPERIMENT}
The results presented in this work are from a single nanocontact STO device with an e-beam patterned $50\times150$~nm$^{2}$ elliptical nanocontact fabricated on top of a 8$\times$26~$\mu$m$^{2}$ pseudo-spin-valve mesa based on Co$_{81}$Fe$_{19}$(20 nm)/Cu(6 nm)/Ni$_{80}$Fe$_{20}$(4.5 nm), as described in Ref.~\onlinecite{mancoff2006apl}. While not shown here, other nanocontacts of varying sizes were also studied as a function of temperature, and gave the same qualitative results.

\begin{figure}[t!]
\includegraphics*[width=0.45\textwidth]{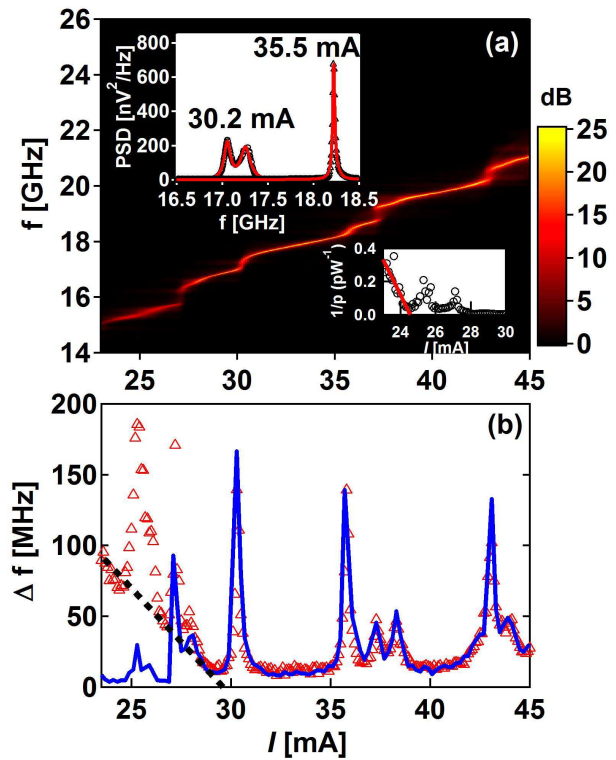}
\caption{(Color online)(a) Two-dimensional power spectral density map of $f$ versus $I$ at a magnetic field of $\mu_0H$=1~T, applied at an angle of $80^{\circ}$ to the film plane. Top inset shows two examples of mode transitions at $I$=30.2~mA and 35.3~mA respectively, where the left spectrum has two clearly resolved Lorentzian peaks, and the right spectrum shows a single broader, asymmetric peak that can still be well fitted by two Lorentzian functions. The bottom inset shows the inverse power $1/p$ vs current and a linear fit (solid line). (b) Experimentally measured (red triangles) and calculated $\Delta f$ (blue solid line). The black dashed line represents a linear fit to linewidth using Eq.(1) for subthreshold currents.}\label{fig:cscan}
\end{figure}
The experimental circuit is similar to that employed in Refs.~\onlinecite{bonetti2009apl} and \onlinecite{muduli2011prb}. The signal generated from the STO was amplified using a broadband +22-dB microwave amplifier, and
detected by a 20 Hz-46 GHz Rohde \& Schwarz FSU46 spectrum analyzer. The measurement was performed in the default mode of the spectrum analyzer,  mode for spectrum analysis, the so-called analyzer mode. We use a resolution bandwidth of 10~MHz and video bandwidth of 10~kHz. The spectra were measured in the frequency range 13-25 GHz with a sweep time of 100~ms. We also average 20 traces resulting in a total measurement time of about 6.4~s.  The dc bias current is fed to the device by a current source through a 0-26 GHz bias tee connected in parallel with the transmission line. The temperature of the sample was varied in the range 300-400~K through use of a heating foil underneath the sample. Each measurement temperature was maintained with a precision of 0.1~K using a thermocouple attached to the bottom of the sample and a software-based PID controller. All measurements were performed in a $\mu_0H$=1~T field applied at an angle of 80$^{\circ}$ w.r.t. the film plane. In this geometry only a propagating spin wave mode~\cite{slonczewski1999jmmm, slavin2005prl, bonetti2010prl,madami2011nn} is excited, and the output power is close to its maximum value.~\cite{bonetti2009apl}

\section{RESULTS}

Figure~\ref{fig:cscan} shows the current ($I$) dependence of the STO frequency at room temperature. In addition to the expected linear blue shift with $I$, a large number of discontinuous jumps and other nonlinearities can be observed. We argue that all these nonlinear features are related to mode transitions, some large, where two distinct peaks can be observed on the spectrum analyzer [the left spectrum in the inset of Fig.~\ref{fig:cscan}(a)], and others small, where only a single peak is observed, though with a significant increase in both nonlinearity and linewidth [the right spectrum in the inset of Fig.~\ref{fig:cscan}(a)]. Similar mode transitions have been observed in the literature~\cite{sankey2005prb,rippard2006prb,krivorotov2007prb} and numerical simulations have reproduced this behavior for in-plane fields.~\cite{berkov2007prb}

\begin{figure}[t!]
\includegraphics*[width=0.45\textwidth]{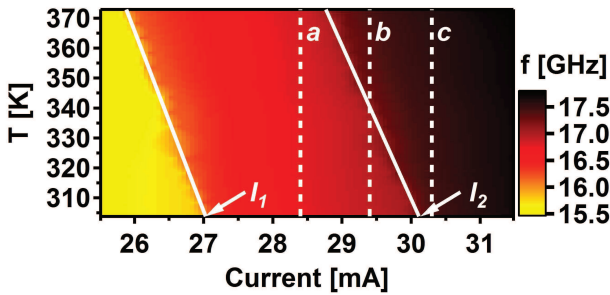}
\caption{(Color online) Map of frequency of the strongest mode versus temperature and bias $I$, showing the mode transition with temperature. The solid lines are linear fits to the threshold current for the mode transitions $I_1$ and $I_2$ versus temperature. The dotted lines are the positions at which the behavior of the linewidth is discussed in Fig.~\ref{fig:lwvstemp}.}\label{fig:freqvstemp}
\end{figure}

The mode transitions have a significant impact on $\Delta f$ vs. $I$, as shown in Fig.~\ref{fig:cscan}~(b). We define $\Delta f$ as the full width at half maximum (FWHM) obtained by fitting a single Lorentzian function. In the case of two modes, we use the linewidth of the strongest mode (the mode with the highest output power). In the subthreshold regime, $\Delta f$ decreases linearly with increasing $I$, which we attribute to the narrowing of the natural ferromagnetic resonance (FMR) linewidth under the influence of the negative damping associated with spin torque.~\cite{kim2008prl1,slavin2009ieeem} At every mode transition position, we also observe a dramatic increase in $\Delta f$ leading to a highly nonlinear dependence on $I$. It is noteworthy that a strong mode transition, and the associated increase in $\Delta f$, can also be observed well inside the subthreshold regime, at about 25~mA. The existence of mode transitions is hence not limited to states of steady precession, as in Ref.~\onlinecite{berkov2007prb}.

\begin{figure}[t!]
\includegraphics*[width=0.45\textwidth]{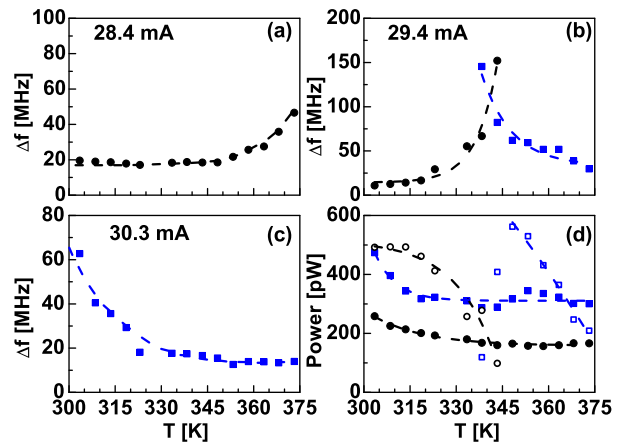}
\caption{(Color online) Measured linewidth versus temperature at (a)~28.4~mA, (b)~29.4~mA, and (c)~30.3~mA. The solid black circles (respectively the solid blue squares) denote the mode excited below (above) $I_{2}$=30~mA at room temperature. (d) Integrated power versus temperature at 28.4~mA (solid black circles), 29.4~mA (open symbols), and 30.3~mA (solid blue squares). The dashed lines serve as visual aids.}\label{fig:lwvstemp}
\end{figure}

In order to show the effect of temperature on mode transitions, we plot a map of measured frequency vs temperature and current, as shown in Fig.~\ref{fig:freqvstemp}. At room temperature these transitions are located at about $I_1$=27~mA and $I_2$=30~mA. As $T$ is increased, both $I_1$ and $I_2$ move to lower values following a linear dependence (the solid lines in Fig.~\ref{fig:freqvstemp}). This $T$ dependence of $I_1$ and $I_2$ has direct consequences for $\Delta f(T)$. To illustrate this, we have chosen three current values, shown by the dashed lines in Fig.~\ref{fig:freqvstemp}, which lie below, on top of, and above the second mode transition. Figures~\ref{fig:lwvstemp} (a)--~\ref{fig:lwvstemp}(c) show $\Delta f$ vs. $T$ at these three currents, which clearly exhibit three dramatically different $T$ dependencies: i) at 28.4~mA, we observe a nonlinear increase of $\Delta f$ with $T$, ii) at 29.4~mA, we observe a nonmonotonic $T$ dependence, and iii) at 30.3~mA we observe an nonlinear \emph{decrease} in $\Delta f$ with $T$. It is quite obvious that none of the measured curves in Fig.~\ref{fig:lwvstemp} follow either a linear or a square-root $T$ dependence, as expected from the theories of thermally induced phase noise.~\cite{sankey2005prb,kim2008prl1,slavin2009ieeem,silva2010ieeemag}

Now we will compare our results with the 
single mode analytical theory.~\cite{kim2008prl1,slavin2009ieeem} According to this theory, $\Delta f$ of a nonlinear oscillator is given by
\begin{eqnarray}
  \Delta f  = & \Gamma_{\rm g}(1-\frac{I}{I_{\rm th}}),  & \mbox{for }\mbox{$I<<I_{\rm th}$} \label{eq:lw1}\\
     = & \Delta f_{\rm L}(1+\nu^2) ,  & \mbox{for }\mbox{$I>>I_{\rm th}$}, \label{eq:lw2}
\end{eqnarray}
where $\Gamma_{\rm g}$ is the natural FMR linewidth,  $I$ the bias current, $I_{\rm th}$ the threshold current, and the nonlinear linewidth amplification is $(1+\nu^2)=1+\left(\frac{p_0N}{\Gamma_p}\right)^2$, where $N=\frac{d\omega}{dp}$ is the nonlinear frequency
shift, and $\Gamma_p$ is the power restoration rate ($\Gamma_p^{-1}$ is the correlation time of the power
fluctuations); $\Delta f_{\rm L}=\Gamma_{\rm g}\frac{kT}{E(p_{0})}$ is the intrinsic thermal linewidth, i.e., the linewidth of a linear ($\nu = 0$) oscillator. Here, ${E(p_{0})}$ is the total energy of the oscillator. Above threshold ($I\gg I_{\rm th}$), the nonlinear amplification of the linewidth is controlled by the ratio of the nonlinear frequency shift $N$ to the power restoration rate $\Gamma_p$.
The reason for this\cite{kim2008prl1,slavin2009ieeem} is that power fluctuations couple to phase
fluctuations through the nonlinear frequency shift $N$, and the linewidth is dominated by phase fluctuations. The linewidth
increases when $N$ is large, so that small power fluctuations give rise to large phase fluctuations, or if $\Gamma_p$
is small, so that power fluctuations remain for a long time during which they affect phase fluctuations.
For the nanocontact under study, the nonlinear damping $Q$ is small\cite{boone2009prb}, and
we can approximate $(1+\nu^2)\approx1+\left(\frac{I}{\Gamma_{\rm g}}\frac{df}{dI}\right)^2$.

We first compare our experimental results for fixed $T$ with theory.\cite{kim2008prl1,slavin2009ieeem} In order to do so,
we need to extract $\Gamma_{\rm g}$. We fit the initial decrease in linewidth with Eq.~(\ref{eq:lw1}), and obtain
$\Gamma_{\rm g}=(500\pm20)$~MHz and $I_{th}=(29\pm1)$~mA, as shown by the dashed line in Fig.~\ref{fig:cscan}~(b). Next, from the measured $f$ vs $I$, we obtain $df/dI$ and directly calculate the nonlinear amplification factor $(1+\nu^2)$, and find from a fit to Eq.~(\ref{eq:lw2}) that $\Delta f_{\rm L}\sim$~67~kHz, for $I>I_{\rm th}$. This value of $\Delta f_{\rm L}$ corresponds to $kT/E(p_{0})\sim1.5\times10^{-4}$. As shown in Fig.~\ref{fig:cscan}~(b), the calculated $\Delta f$ shows very good agreement with the experimentally measured linewidth, and also reproduces the dramatic increase in $\Delta f$ which occurs around each mode transition. The agreement indicates that the nonlinear amplification of the linewidth is
controlled by the nonlinear frequency shift $N\propto df/dI$, while the power restoration rate $\Gamma_p$ is
constant. The agreement is lost for $I<$27~mA, as expected for currents below threshold.~\cite{kim2008prl1,slavin2009ieeem} We have also used the inverse power method~\cite{tiberkevich2007apl} to determine the threshold current as shown in the inset of Fig.~\ref{fig:cscan}~(a). A fit of this data for current below 25 mA is shown by the solid line. From this fit it appears as if the STO is close to auto-oscillation already at about 24.5~mA, but gets interrupted by one or more subthreshold mode transitions. It is only at about 27-28 mA that robust auto-oscillation begins.

\begin{figure}[t!]
\includegraphics*[width=0.45\textwidth]{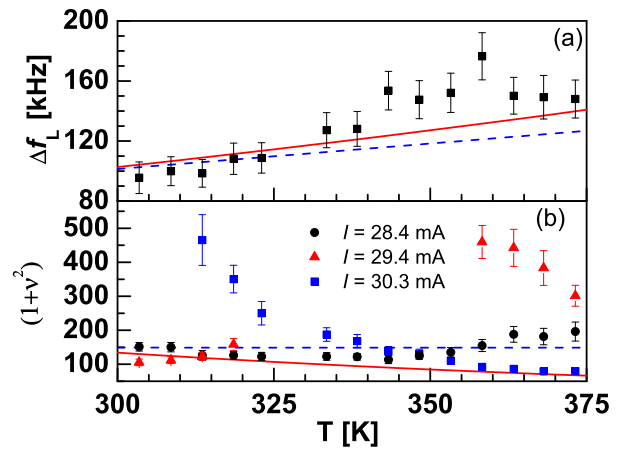}
\caption{ Temperature dependence of (a) extracted linear contribution of linewidth $\Delta f_{\rm L}$ (solid symbols) and (b) the nonlinear amplification, $(1+\nu^2)$ (solid and open symbols). The solid red lines are calculation based with inclusion of temperature dependence of $M_{\rm s}$, where as the dashed blue lines are calculation assuming no temperature dependence of $M_{\rm s}$.}\label{fig:LlwNLvstemp}
\end{figure}

Next, we want to compare the temperature dependence (at fixed $I$) of $\Delta f_{\rm L}$ and $(1+\nu^2)$ as obtained
from the experiment with theoretical predictions\cite{kim2008prl1,slavin2009ieeem}. According to the
theory, $\Delta f_{\rm L}(T)$ should be proportional to $T$, since it is the linewidth of a linear oscillator
in contact with a thermal bath, while $(1+\nu^2)$ has a monotonic temperature dependence. Using the agreement between the calculated and measured linewidths in Fig.~\ref{fig:cscan}, we can now extract $\Delta f_{\rm L}$ and its temperature dependence, as shown in Fig.~\ref{fig:LlwNLvstemp}(a). Since the determination of $(1+\nu^2)$ is more accurate in regions between mode transitions, i.e., where $df/dI$ do not diverge, we use the average value of $\Delta f_{\rm L}$ for 30.5~mA$<I<31.5$~mA, which excludes any mode transitions and is above threshold at all temperatures. A linear increase in $\Delta f_{\rm L}$ with $T$ is observed. The solid and dashed lines are calculations based on the classical quasi-Hamiltonian formalism for spin waves,~\cite{slavin2005ieeem,slavin2008ieeem,slavin2009ieeem,kim2008prl,kim2008prl1} which shows reasonable agreement with the experiment and also predicts a linear behavior similar to experiment even with the inclusion of the temperature dependence of $M_0$ in the calculation (red solid line). This calculation assumes single mode excitation but considers the nonuniform nature of propagating spin waves by "exchange normalization" of magnetic field, and normalization of volume under the nanocontact.~\cite{slavin2009ieeem} The parameters used are similar to those of Ref.~\onlinecite{bonetti2010prl}, the electron gyromagnetic factor: $\gamma=1.76\times10^{7}$ rad/Oe, saturation magnetization: $M_0(300 K)=640$ emu/cm$^{3}$, Gilbert damping parameter: $\alpha_G=0.01$, dimensionless spin-polarization efficiency: $\epsilon=0.2$, the exchange length: $\lambda_{\rm ex}=5$~nm, and $(I/I_{\rm th})_{300 K}=5$. The effective volume $V_{\rm eff}$ is assumed to be 1.5 times that of the volume under the nanocontact. We use $\Gamma_{\rm g}=(500\pm 20)$~MHz, as determined from the experiment. Calculation also predicts $\Gamma_{\rm g}=500$~MHz for our experimental geometry. We note that the agreement with $\Delta f_{\rm L}$ with $T$ was obtained only when $I/I_{\rm th}>5$. We attribute this to the fact that the analytical Eq.~(\ref{eq:lw2}) is an asymptotic equation that is valid only for $I>>I_{\rm th}$.~\cite{kim2008prl1} Basically we treated
$I/I_{\rm th}$ as a fitting parameter, with the other parameters kept fixed at their reasonable values, since the precise values of these parameters are a bit uncertain.

\begin{figure}[t!]
\includegraphics*[width=0.45\textwidth]{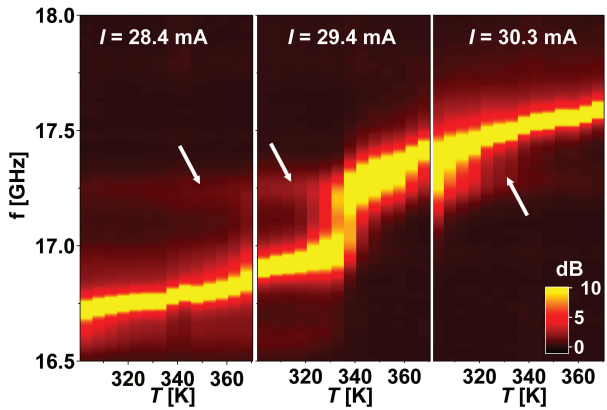}
\caption{ Map of power (dB) vs frequency ($f$) and temperature $T$ for the three example current values of 28.4~mA, 29.4~mA, and 30.3~mA. The arrows indicate the presence of additional modes, the  amplitude of which depends on temperature.}\label{fig:spectmap}
\end{figure}

In Fig.~\ref{fig:LlwNLvstemp}(b) we show the behavior of measured $(1+\nu^2)$ vs $T$ (symbols) for the three current values of 28.4~mA, 29.4~mA and 30.3~mA along with the calculated behavior for $I/I_{\rm th}=5$ (solid and dashed lines). The experimental behavior of $(1+\nu^2)$ vs $T$ is dramatically different for the three cases but very similar to the behavior of the linewidth
as a function of $T$ shown in Fig~\ref{fig:lwvstemp}. In contrast, the calculations of single-mode theory predict a monotonic decrease of $(1+\nu^2)$ with $T$ when the temperature dependence of  $M_{\rm s}$ is included. Hence the calculations agree with the experiment only for a limited range of temperature and when the STO is far from the mode transition. For example, at 28.4~mA (30.3~mA), $(1+\nu^2)$  is enhanced at higher (lower) temperature, which is close to the mode transition. Detail examination show that this enhancement occurs when two modes are observed. This can be clearly seen in Fig.~\ref{fig:spectmap} which shows the measured power {\it vs} frequency ($f$) and temperature $T$
for the three current values. These spectra show the presence of  an  additional mode (as shown by the arrows) for all there currents. The temperature dependence of the amplitude of this additional mode has a clear correlation with the behavior of $(1+\nu^2)$ vs $T$ in Fig.~\ref{fig:LlwNLvstemp}(b). For example, the amplitude of second mode increases (decreases) with temperature at 28.4~mA (30.3~mA). Thus our results indicate that when two modes are observed, the experimental $(1+\nu^2)$ is enhanced compared to the prediction of single mode calculation.

\section{DISCUSSION}

We will now discuss the mechanism for the anomalous temperature dependence of the linewidth. The basic assumption is
that in the presence of two mode,  mode coupling near a transition can lead to an increase in the linewidth.
The starting point is a set of coupled equations for the complex amplitudes $c_i$, $i=1,2$ of the
time-dependence of the modes,~\cite{muduli2012prl}
\begin{widetext}
\begin{eqnarray}
\frac{dc_1}{dt}+i\omega_1\left( p_1,p_2\right)c_1
+\left[\Gamma_+\left(p_1,p_2\right)-\Gamma_-\left(p_1,p_2\right)\right]c_1
-ke^{i\varphi}c_2 &=&0\nonumber\\
\frac{dc_2}{dt}+i\omega_2\left( p_1,p_2\right)c_2
+\left[\Gamma_+\left(p_1,p_2\right)-\Gamma_-\left(p_1,p_2\right)\right]c_2
-ke^{i\varphi}c_1 &=&0.\label{eq:coupled_eqn_1}
\end{eqnarray}
\end{widetext}
Here, $\Gamma_+$ and $\Gamma_-$ are the positive and negative damping, and $\omega_1$ and $\omega_2$ the mode-frequencies;
we have indicated the dependence of $\omega_i$, $\Gamma_+$, and $\Gamma_-$ on the mode powers $p_1$ and $p_2$. The equations
contain a linear coupling term with complex amplitude $ke^{i\varphi}$, with $k$ real and $k\geq0$. This term is not allowed on short
time scales if $\omega_1\not=\omega_2$. Here, however, we are interested in behavior over times much larger than the
time-scale of the periods of the modes or of thermal fluctuations. In that case, the coupling mediated through the linear coupling
describes processes in which one mode can decay into the other through intermediate states and energy that is absorbed or
released into other magnetic modes or a thermal reservoir. Such a process becomes more likely as the mode frequencies approach
each other, with a concomitant increase in $k$. The experiments show a significant current and temperature dependence of the main mode frequency. Therefore, the linear mode coupling also has
a strong current and temperature dependence, $k=k(I,T)$ and $\varphi=\varphi(I,T)$. In particular, the magnitude of
$k$ has maxima at currents and temperatures at which mode transitions occur. We will see that this coupling plays a key role.

We now make some simplifying assumptions. First, we assume that the mode frequency $\omega_i$
only depends on $p_i$ and not
on $p_j$, $j\not=i$. Next, we assume that the system is close to, but above, threshold (recall that the threshold current is about 27 mA and the
relevant current values are around 30 mA). We then expand Eq.~(\ref{eq:coupled_eqn_1}) near 
$c_i=0$ and write the equations in terms of power amplitude and phase,
$c_i=\frac{Q_i}{\sqrt{(\omega_i}}e^{-i\left(\omega_{i,0}t-\varphi_i\right)},
$
where $\omega_{i,0}$ is the threshold mode frequency. This leads to the following equations for the time dependence of the
amplitudes $Q_i$ and phases $\varphi_i$:
\begin{subequations}
\label{eq:coupled_eqns_2}
\begin{eqnarray}
\frac{dQ_1}{dt} & = & \Gamma_g \left( I/I_{\rm th}-1\right) Q_1 -\left( \overline Q Q_1^2+\overline PQ_2^2\right) Q_1\nonumber\\
&&+kQ_2\sqrt{ \frac{\omega_{1,0}}{\omega_{2,0}}} \cos(\varphi-\varphi_2+\varphi_1) \label{eq:coupled_eqn_2a}\\
\frac{dQ_2}{dt} & = & \Gamma_g \left(I/I_{\rm th}-1\right) Q_2 -\left( \overline Q Q_2^2+\overline PQ_1^2\right) Q_2\nonumber\\
&&+kQ_1 \sqrt{\frac{\omega_{2,0}}{\omega_{1,0}}} \cos(\varphi-\varphi_2-\varphi_1) \label{eq:coupled_eqn_2b}\\
\frac{d\varphi_1}{dt} & = & -N_1Q_1^2+k\frac{Q_2}{Q_1}\sqrt{ \frac{\omega_{1,0}}{\omega_{2,0}}}\sin(\varphi+\varphi_2-\varphi_1)
\label{eq:coupled_eqn_2c}\\
\frac{d\varphi_2}{dt} & = & -N_2Q_2^2+k\frac{Q_1}{Q_2}\sqrt{\frac{\omega_{2,0}}{\omega_{1,0}}}\sin(\varphi-\varphi_2+\varphi_1).
\label{eq:coupled_eqn_2d}
\end{eqnarray}
\end{subequations}
Here, $N_i$ is the nonlinear frequency shift, $I_{\rm th}$ the threshold current, and $\overline Q$ and $\overline P$ the diagonal
and off-diagonal nonlinear damping coefficients, respectively. We will for simplicity assume that $N_1=N_2=N$.

Next, we introduce the transformations\cite{vanderSande} $Q_1=\sqrt{p}\cos\left(\frac{\theta+\pi/2}{2}\right)$ and
$Q_2=\sqrt{p}\sin\left(\frac{\theta+\pi/2}{2}\right)$, where $p$ is the total power in the two modes. These transformations
recast the description of the mode amplitudes $Q_i$ in terms of $p$ and $\theta$, where $p$ is total power and $\theta$
describes how the power is distributed between the two modes.
Inserting
these in Eqs.~(\ref{eq:coupled_eqn_2a}) and (\ref{eq:coupled_eqn_2b}) and assuming
that the {\em average} power $p_0$ is stationary and writing $p=p_0+\delta p$, with $dp_0/dt=0$ and $\delta p$ the power fluctuations, we obtain the following
linearized equation for the power (linearized in power fluctuations about the average power $p_0$):
\begin{widetext}
\begin{eqnarray}
\frac{d(p_0+\delta p)}{dt} & = & 2\left( I/I_{\rm th}-1\right)\Gamma_g\left(p_0+\delta p\right)
-2\overline Q\left(p_0^2+2p_0\delta p\right)-\left(\overline P-\overline Q\right)\cos^2\theta\left(p_0^2+2p_0\delta p\right)\nonumber\\
&&+{k}\left(p_0+\delta p\right)\cos\theta\left[\sqrt{\frac{\omega_{1,0}}{\omega_{2,0}}}\cos\left(\varphi+\psi\right)
+\sqrt{\frac{\omega_{2,0}}{\omega_{1,0}}}\cos\left(\varphi-\psi\right)\right],
\label{eq:power_fluct_1}
\end{eqnarray}
where $\psi=\varphi_2-\varphi_1$, with a time evolution given by
\begin{eqnarray}
\frac{d\psi}{dt} & = & -Np_0\sin\theta-N\delta p\sin\theta
+k\frac{1-\sin\theta}{\cos\theta}\sqrt{\frac{\omega_{2,0}}{\omega_{1,0}}}\sin(\varphi-\psi)
-k\frac{1+\sin\theta}{\cos\theta}\sqrt{\frac{\omega_{1,0}}{\omega_{2,0}}}\sin(\varphi+\psi)
\label{eq:phase_fluct_1}
\end{eqnarray}
\end{widetext}
Ignoring fluctuations for the moment, and keeping
in mind that the power $p_0$ is constant, the equations~(\ref{eq:power_fluct_1}) and (\ref{eq:phase_fluct_1}) describe a
two-dimensional dynamically driven system in $(\theta,\psi)$-space.  The system under consideration here has, far away from a mode transition so that $k\approx0$, a single
stable fixed point $\theta=-\pi/2$ ($\theta=\pi/2$) with all power in mode $\omega_1$ ($\omega_2$) well below (above) the mode transition. Near or at the mode transition, the system may have stable fixed points, unstable fixed points, or limit cycles. In either case we will assume that the experimental linewidth arises from fluctuations in the {\em total} power and phase difference, and we
will therefore ignore fluctuations in $\theta$. By enforcing the stationarity condition $dp_0/dt=0$ we obtain from Eq.~(\ref{eq:power_fluct_1})
\begin{widetext}
\begin{equation}
2\left(I/I_{\rm th}-1\right)\Gamma_g-2\overline Qp_0-\left(\overline P-\overline Q\right)p_0\cos^2\theta
+kp_0\cos\theta\left[\sqrt{\frac{\omega_{1,0}}{\omega_{2,0}}}\langle\cos(\varphi+\psi)\rangle
+\sqrt{\frac{\omega_{2,0}}{\omega_{1,0}}}\langle\cos(\varphi-\psi)\rangle\right]=0,
\label{eq:constant_p}
\end{equation}
\end{widetext}
where $\langle\ldots\rangle$ denotes a suitable time-average over times long compared to the time scale of fluctuations ({\em e.g.,} a limit
cycle).
Inserting this into Eq.~(\ref{eq:power_fluct_1}), and separating $\psi$ into a regular part $\Psi$, describing the
slow time evolution of the phase difference of the two modes, and fluctuations
$\delta\psi$, $\psi=\Psi+\delta\psi$,
and replacing $\cos\psi$ ($\sin\psi$) with $\langle\cos\Psi\rangle$ ($\langle\sin\Psi\rangle$) we obtain
the following linearized equations relating the fluctuations in power and phase angle difference:
\begin{widetext}
\begin{eqnarray}
\frac{d\delta p}{dt} & = & 2\left(I/I_{\rm th}-1\right)\Gamma_g\delta p-4\overline Qp_0\delta p+2\left(\overline P-\overline Q\right)
\cos^2\theta\delta p
+k\delta p\cos\theta\left[ \sqrt{\frac{\omega_{1,0}}{\omega_{2,0}}}\langle\cos(\varphi+\Psi)\rangle+\sqrt{\frac{\omega_{2,0}} {\omega_{1,0}}}\langle\cos(\varphi-\Psi)\rangle\right]\nonumber\\
&&-kp_0\cos\theta\delta\psi           \left[ \sqrt{\frac{\omega_{1,0}}{\omega_{2,0}}}\langle\sin(\varphi+\Psi)\rangle
-\sqrt{\frac{\omega_{2,0}} {\omega_{1,0}}}\langle\sin(\varphi-\Psi)\rangle\right],
\label{eq:linear_fluct_2}
\end{eqnarray}
\end{widetext}
and
\begin{eqnarray}
\frac{d\delta\psi}{dt}  &=&  -N\delta p\sin\theta\nonumber\\
&&-k\delta\psi\frac{1-\sin\theta}{\cos\theta} \sqrt{\frac{\omega_{2,0}}{\omega_{1,0}}}\langle\cos(\varphi-\Psi)\rangle\nonumber\\
&&-k\delta\psi\frac{1+\sin\theta}{\cos\theta} \sqrt{\frac{\omega_{1,0}}{\omega_{2,0}}}\langle\cos(\varphi+\Psi)\rangle
\label{eq:fluct_3}
\end{eqnarray}
with $\Psi$ satisfying
\begin{eqnarray}
\frac{d\Psi}{dt} & = & -Np_0\sin\theta\nonumber\\
&&+k\frac{1-\sin\theta}{\cos\theta}\sqrt{\frac{\omega_{2,0}}{\omega_{1,0}}}\sin(\varphi-\Psi)\nonumber\\
&&-k\frac{1+\sin\theta}{\cos\theta}\sqrt{\frac{\omega_{1,0}}{\omega_{2,0}}}\sin(\varphi+\Psi).
\label{eq:Phi_eqn}
\end{eqnarray}
We pause for a moment to note that Eqs.~(\ref{eq:linear_fluct_2}) to (\ref{eq:Phi_eqn}) restricted to a single mode
($k=0$ and $\cos\theta=0$)
are precisely the results of Kim, Slavin, and Tiberkevich\cite{slavin2005ieeem,slavin2008ieeem,slavin2009ieeem,kim2008prl},
with $\Gamma_p=-\left(I/I_{\rm th}-1\right)\Gamma_g+2\overline Qp$, describing the power fluctuations in the oscillator, and
how the power fluctuations couple to the phase fluctuations through the nonlinear frequency shift $N$. It is of course
this latter coupling that gives rise to the enhanced linewidth through the enhanced phase fluctuations. As we
noted earlier, in the low-temperature limit,
applicable here, the single-oscillator linewidth enhancement is described by the ratio of the nonlinear frequency shift $N$ to the
power restoration rate $\Gamma_p$: Power fluctuations couple to the nonlinear frequency shift, and the longer the decay time
of power fluctuations is ({\em i.e.,} smaller $\Gamma_p$), the more power fluctuations can affect phase fluctuations. For the system
under consideration here, Eqs.~(\ref{eq:linear_fluct_2}) and (\ref{eq:fluct_3}) show that the mode-coupling $k$ leads to
additional coupling between power and phase fluctuations. In general, the solutions to these equations, especially in the presence
of thermal fluctuations, are complicated. We can, however, gain some insight by assuming that $\omega_{1,0}\approx\omega_{2,0}$ with
$\omega_{2,0} > \omega_{1,0}$
and consider the system far from a mode transition so that $k$ is small and $\theta=-\pi/2+\delta$, with
$\delta\ll1$, and $\varphi$ small and negative. For the nanocontact STOs, the nonlinear amplification $N$ is large, and the nonlinear damping small. First, with
$N$ large and $k$ small, we can neglect the terms in $\delta\psi$ on the right-hand side of Eq.~(\ref{eq:fluct_3}). This means
that power amplitude fluctuations couple to phase fluctuations through $N$ just as for the single oscillator. It follows that if the power
amplitude fluctuations are enhanced or prolonged by the mode coupling so that $\delta p$ is enhanced or $\Gamma_p$ reduced
by the mode coupling,
Second, for $N$ large and the nonlinear damping small,  at the fixed point $\theta\approx-\pi/2$ we have
$\cos(\varphi-\Psi)\approx 0$ and $\sin(\varphi-\Psi)\approx -1$. If we neglect the terms in $\delta\psi$ on the right-hand
side of Eq.~(\ref{eq:linear_fluct_2}), the net effect under these assumptions is to change the power restoration rate  $\Gamma_p\to\Gamma_p-k\cos(\theta)\sin(|\varphi|)$ with a concomitant enhancement
of the nonlinear amplification and the linewidth as the coupling term $\nu$  between power amplitude and phase fluctuations is given by $\nu=Np_0/\Gamma_p$ in single mode theory.\cite{slavin2009ieeem} This explains qualitatively why the observed dependence of the linewidth on temperature in general does not agree with the theoretical expression\cite{slavin2009ieeem} (Fig.~\ref{fig:LlwNLvstemp}). In the latter, the temperature
dependence is driven by the stochastic thermal noise. In contrast, the experimentally
determined nonlinear amplification contains a modified power restoration rate that includes
the temperature (and current) dependence of $k$ (and $\varphi$). %

\section{CONCLUSIONS}
In conclusion, we have shown that the behavior of spin torque oscillator linewidths is to a large extent determined by nonlinearities arising from a number of mode transitions. The mode transitions are observed at increasing current at fixed temperature, or at increasing temperature
at fixed current. Near the mode transitions, the linewidth increases substantially. Nevertheless, both the current and temperature
dependence of the linewidth are well described by the analytical single-oscillator theory using the nonlinear
amplification extracted from experimental data. In contrast, the temperature dependence of the linewidth near the mode transitions does not
agree well with the single-oscillator analytical theory if the nonlinear amplification is calculated directly from the theory. The experimental
data showed the presence of an additional mode where the nonlinear amplification is enhanced near the mode transitions. We have
argued that a temperature-dependent mode coupling leads to reduction of the power restorations rate, and therefore
an enhancement of the nonlinear amplification and of the linewidth, and that this at least qualitatively explains the anomalous temperature dependence of the linewidth near the mode transitions. These results are important for the understanding of linewidth in spin torque oscillators.

\section*{ACKNOWLEDGEMENTS}
We thank Fred~Mancoff at Everspin Technologies, USA for providing the samples used in this work. We also thank S.~Bonetti and Niels de Vreede for assistance in experiments and useful discussions. Support from the Swedish Foundation for Strategic Research (SSF), the Swedish Research Council (VR), and the G\"{o}ran Gustafsson Foundation are gratefully acknowledged. Knut and Alice Wallenberg foundation (KAW), is acknowledged for funding of the equipment used for measurements presented here. P. M. acknowledges Swedish Research Council (VR) for the "Junior Researchers Project Grant". J.~\AA .  is a Royal Swedish Academy of Sciences Research Fellow supported by a grant from the Knut and Alice Wallenberg Foundation. Argonne National Laboratory is operated under Contract No. DE-AC02-06CH11357 by UChicago Argonne, LLC.

\end{document}